# WHITE PAPER 2019

# IOT EVOLUTION TOWARDS A SUPER-CONNECTED WORLD

Internet of Things (IoT) is one of the key components of Digital Transformation, along with big data and analytics. IoT, together with the cloud, big data, analytics, machine learning (ML), and deep ML, can help create numerous possibilities and new opportunities. These possibilities will impact our daily lives substantially and open new business models for consumers and enterprises where the number of connected IoT devices could go up to 50 Billion by 2022. The ecosystem of IoT consists mainly of sensors/devices layer, connectivity layer and IoT platform. The main value of IoT is in creating use cases for efficiency, monitoring and management of the things/devices. IoT connects the things through the Internet to the IoT platform which equipped with device management and with the possibility of creating new use cases along with data analytics and ML that provide 360 view through data insight.[1]

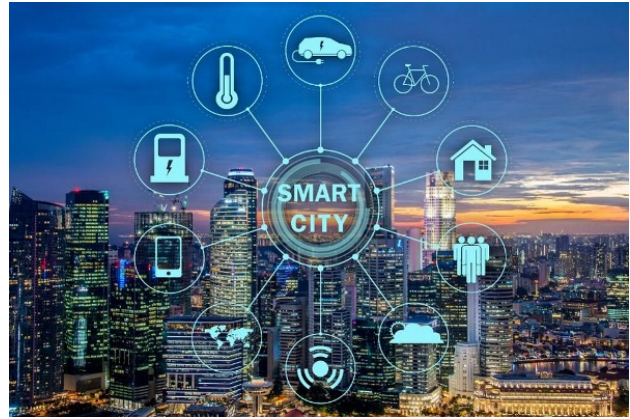

**Author: Ayman Elnashar**[*]

---

[*] Head and VP – Technology Planning – ICT & Cloud at du of Emirates Integrated Telecommunications Co. (EITC)

# 1. INTRODUCTION

In order to set the foundation for the IoT evolution, a traffic profile analysis for a commercial LTE/3G/2G network for different types of devices, such as smartphone, feature phone, and Machine to Machine (M2M)/IoT Modules, is illustrated in Figure 1.[2] The traffic profiles compare the mobile broadband traffic versus M2M/IoT traffic. The analyzed network deployed 2G, 3G, and 4G with almost 100% coverage. In this mobile broadband case, increasing 4G devices in the network is healthy for Average revenue per user (ARPU) and maximum utilization of the data network. On the other side, the network does not have any cellular IoT type technology (Low Power Wide Coverage network such as NB-IOT or eMTC), and therefore, the M2M/IoT devices are using the normal 2G/3G/4G connectivity. Most of the M2M/IoT modules observed in the network are POS machine, industrial, and manufacturing M2M modules.[2]

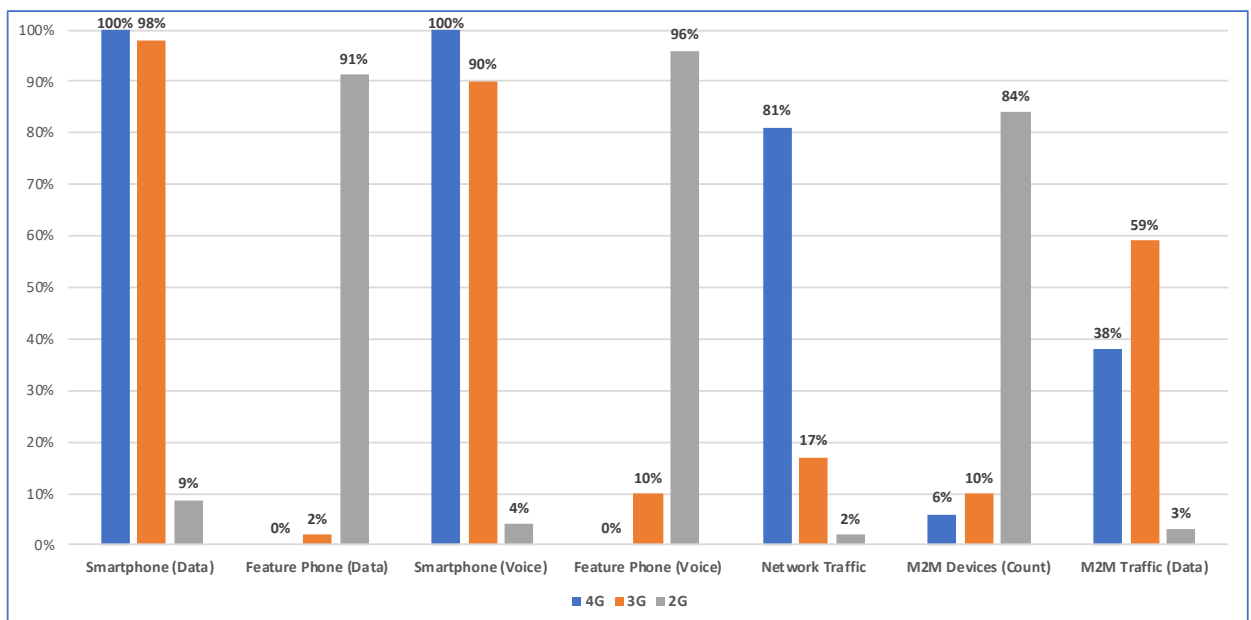

Figure 1 Case Study for Broadband Traffic Analysis

The following summarizes the observations from the above use case:

1- 98% of the network traffic is coming from 4G/3G Smartphones while 2G traffic is only 2% of the total network traffic.
2- The contribution of M2M and IoT modules to overall data traffic is only 0.3% while the penetration of M2M/IoT devices is less than 4%. This is going to be changed dramatically after the introduction of cellular IoT devices such as NB-IoT and LTE-M devices, with a massive number of devices and very low traffic.
3- The 4G and 3G traffic of M2M modules is about 97% while their penetration is only 16% of the M2M devices. Again, this is mainly because the deployed M2M devices are mainly industrial M2M use cases.

Introduction of NB-IoT and LTE-M is mandatory to handle the IoT use cases, especially with low power, extended coverage, and low-cost modules. The service providers (SP) should adopt and promote a clear IoT/M2M strategy in order to be able to cater for all use cases and to efficiently manage the spectrum resources. The following options need to be analyzed based on the SP strategy:

**Option 1**: Shut down of 2G Network: in this case, the SP need to migrate 80% of exiting M2M devices to 3G/4G or to NB-IoT and LTE-M, based on the use case requirements.

**Option 2**: Shut down of 3G network: in this case, the SP need to migrate only 10% of M2M devices to 4G CAT1 or LTE-M. This depends on the use case and availability of M2M devices in 4G or LTE-M.

**Option 3**: Shut down of 2G and 3G networks (long term strategy): in this case, 94% of M2M/IoT devices need to be migrated to LTE/LTE-M or NB-IoT, based on the use case and device availability.

The SP need to identify the right strategy for cellular IoT introduction; some operators adopt only NB-IoT while other operators adopt LTE-M or both technologies. This depends on the targeted use cases and operator strategy. Some other operators (early LPWA adopters) have already deployed non 3GPP technologies such as LoRa and SigFox. However, they may deploy LTE-M or LTE CAT1 for higher throughput devices or even deploy NB-IoT at a later stage. This depends on the region, targeted use cases, maturity of the technologies, availability of the spectrum, and many other factors. The IoT use cases are mainly derived by large enterprises, and accordingly, SP need to be smart and adopt the right strategy to capture all opportunities. This paper will provide a proposed strategy for SP in section 3.

## 2.  CELLULAR 3GPP IOT TECHNOLOGIES

The Cellular 3GPP IoT standard has evolved to address the following challenges:

- Low power devices with up to 10 years on battery for use cases such as smart parking.
- Increased coverage for IoT network, up to 20 dB coverage to cater for use cases such as a smart meter installed in a basement.
- Massive IoT device numbers support up to 100k devices per base station to cater for massive IoT use cases such as environmental sensors.

As a result, 3GPP has introduced two technologies: NB-IoT and eMTC at 3GPP Rel 13 as a software upgrade to existing LTE networks. Therefore, 3GPP cellular technologies (2G/3G/LTE/NB-IoT/eMTC and later 5G) can address any IoT applications thanks to:
- Wide area coverage of existing network
- Low cost per device
- Connectivity by large scale of devices
- Functionality of UE mobility
- Ultra-reliable low latency (URLL) applications with 5G.

Table 1 illustrates the different IoT applications and their requirements.[2]

Table 1: Typical IoT use case requirements

| Applications | Battery Life <2yrs/ Mid/ >10 (Long) | Coverage | Latency | Mobility | Data rate |
|---|---|---|---|---|---|
| Utility meters | Long | Deep indoor coverage (Extreme coverage) | High | Stationary | Low ~ 100bps to some kpbs |
| Payment transactions (POS terminals at retail establishments and kiosks) | Wall powered. | Outdoor/indoor, deep coverage | Mid to high | Stationary | Low ~some kbps |
| Tracking of people, pets, vehicles and assets | Long | Outdoors / indoors (extreme coverage) | Low/Mid | Mobile/ Nomadic | Low ~ up to 100kbps |
| Wearable | Same as smart phone | Normal coverage | Low | As LTE | High |
| Home alarm panels with and without voice | High/Mid | Normal to extended | Mid | Stationary | Low/high depending on voice/video |
| Automotive | On car battery | Normal to extended coverage | Mid to low or very low | Mobility | From low to high |
| Industrial control | Wall powered | Normal | Low to extremely low | Stationary | Might be large |

The cellular network evolution towards 5G has started and will continue until 2022 and beyond. The IoT defines the way for intelligently connected devices and systems to leverage and exchange data between small devices and sensors in machines and objects. IoT concepts and working models have started to spread rapidly, which is expected to provide a new dimension for services that improve the quality of a consumer's life and the productivity of enterprises. The IoT effort started from the concept of M2M solutions to use wireless networks to connect devices to each other and through the Internet in order to deliver services that meet the needs of a wide range of industries.

IoT must deal with the numerous challenges of a massive number of cheap devices providing low energy consumption connected in a wider range, referred to as Low Power Wide Area (LPWA). Therefore, IoT is typically classified into 1.) IoT connectivity in unlicensed spectrum; and 2.) cellular IoT in licensed spectrum. Many technologies are emerging to deal with these two categories, including:

- **Unlicensed Networks** – For short range, we have technologies such as Bluetooth Low Energy, Wi-Fi, IEEE802.11ah, IEEE802.15.4, ZigBee, and Z-wave. For the long range, there are Sigfox, Weightless, OnRamp, and LoRa.

- **Cellular IoT in licensed spectrum** – Which includes LTE evolution for Machine Type Communication (MTC), Narrow-band IoT, and the GSM evolution of IoT referred to as Extended-coverage GSM (EC-GSM).

Figure 2 demonstrates the forecast for different LPWA technologies.[4] As demonstrated in Figure 2, the 3GPP based technologies are excepted to dominate the market, although they started after the unlicensed technologies. The main reason is the global ecosystem and open standard supported by major telecom giants' vendors, chipset, and service providers. The lack of global standardization for LPWA in the unlicensed bands and interference are the major bottleneck to deploying critical IoT applications in these bands, such as utilities and industrial IoT (IIoT).

Also, Figure 2 indicates a steadily growing market of annual connected devices, increasing 20% year over year. More use cases will be realized when Cellular IoT Networks become globally available. LPWA Cellular IoT started in 2018 and is expected to increase the global adaptation within 2019/2020 for specific use cases such as smart meters, fire alarm systems, smart parking, wearables, and smart tracking/logistics use cases.

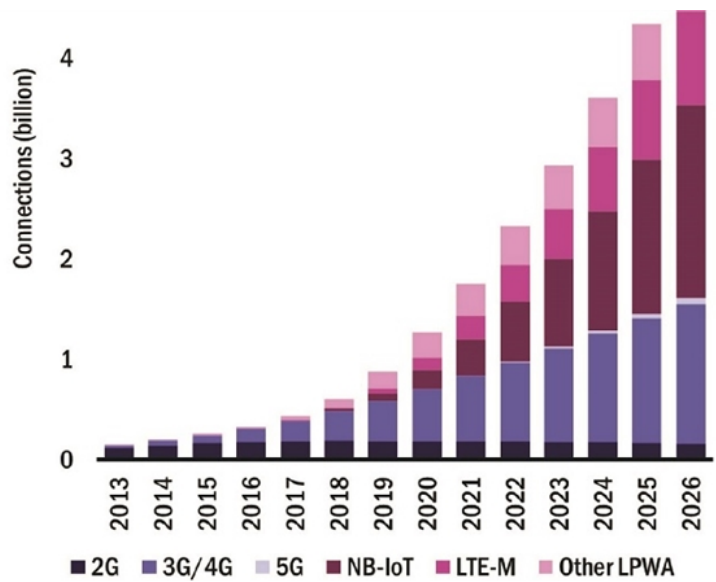

Figure 2: LPWA connections forecast[4]

Use cases for machine-type communications are evolving very rapidly. There has been enormous interest in integrating connectivity solutions with sensors, actuators, meters (water, gas, electric, or parking), cars, appliances, etc. IoT consists of a number of networks that may have different design objectives. For example, some networks only intend to cover the local area (e.g. one single home), whereas some networks offer wide-area coverage. The latter case is being addressed in the 3GPP. Recognizing the importance of IoT, 3GPP has introduced a number of key features for IoT in its latest release: Rel-13. EC-GSM-IoT and LTE-MTC aim to enhance existing GSM and LTE networks, respectively, for better serving IoT use cases. Coverage extension, UE complexity reduction, long battery lifetime, and backward compatibility are common objectives. A third track, NB-IoT, shares these objectives as well. In addition, NB-IoT aims to offer deployment flexibility, allowing an operator to introduce NB-IoT using a small portion of its existing available spectrum. NB-IoT is designed to mainly target ultra-low-end IoT applications. Refer to https://www.gsma.com/iot/mobile-iot-commercial-launches/ for commercially deployed NB-IoT and LTE-M networks worldwide. Table 2 provides a comparison between cellular IoT technologies up to 3GPP Rel 14 and is summarized as follows:[3,5]

- **LTE CAT-1** - It's an excellent option for IoT applications that require a browser interface or voice. The major attraction is that it's already standardized, and more importantly, it's simple to move into CAT-1 devices. It is expected that CAT-1 devices can replace 3G M2M devices. Typical examples for LTE CAT1 are smart meter concentrators, smart fire alarm systems, and smart buildings. The adoption of CAT 1 instead of fully-fledged CAT 4 LTE is due to simplicity of design, low TP for specific use cases, being replaceable for 3G M2M devices, and lower cost compared to LTE CAT 4 or even 3G. In some scenarios, a CAT 1 modem may come as LTE only or with a 3G/2G fall back option for better availability.

- **LTE CAT-M:** This is often viewed as the second generation of LTE chips built for IoT applications. It offers cost and power consumption reduction by limiting the maximum system bandwidth to 1.4 MHz (as opposed to 20 MHz for Cat-1). Cat-M is really targeting LPWAN applications like smart metering where only a small amount of data transfer is required. Also, wearables with VoLTE option for voice may adopt CAT-M. The CAT-M is compatible with the existing LTE network. However, a new software upgrade for the access network is needed and, possibly, a license in the core network. Also, proper network planning is needed to deploy an LTE-M layer on top of the existing LTE network. The Infra vendor will try to change the license model for LTE network to be per device or per traffic volume. It is very important for SP to adopt the right licensing model. The best strategy here is a *pay as you grow* model, which limits the license to the number of devices initially without any cap on the traffic as LTE-M devices are expected to consume higher traffic.

- **NB-IoT:** This can be treated as a new radio interface with extreme low bandwidth requirements, which provides high efficiency of deployment and very low device complexity. NB-IoT inherited the same protocol as LTE to handle network signaling with a smaller number of features. With Rel-14 enhancement, eNB-IoT upgrades the supported data rates to extend the usage cases for NB-IoT but still keeps the competitive UE complexity as Rel-13 NB-IoT. NB-IoT is the real 3GPP evolution that addresses the LPWA use cases and competes with LoRa and Sigfox and other unlicensed LPWA technologies. NB-IoT fits use cases such as smart parking, smart meter, smart environment, etc.[5]

Table 2: Comparison between 3GPP Cellular IoT Technologies

| Criterion | Cat. 1 (Rel. 8+) | Cat. M1 (Rel. 13) | Cat. NB1 (Rel. 13) | FeMTC (Rel. 14) | eNB-IOT (Rel. 14) |
|---|---|---|---|---|---|
| Bandwidth | 20 MHz | 1.4 MHz | 180 kHz | Up to 5 MHz (CE Mode A and B for PDSCH and A only for PUSCH) | 180 kHz |
| Deployments/ HD-FDD | LTE channel / No HD-FDD | Standalone, in LTE channel / HD-FDD preferred | Standalone, in LTE channel, LTE guard bands, HD-FDD | Standalone, in LTE channel / HD-FDD, FD-FDD, TDD | Standalone, in LTE channel, LTE guard bands, HD-FDD preferred |
| MOP | 23dBm | 23dBm/ 20dBm | 23dBm/ 20dBm | 23dBm / 20dBm | 23dBm/ 20dBm/ 14dBm |
| Rx ant / layers | 2/1/ | 1/1 | 1/1 | 1/1 | 1/1 |
| Coverage, MCL | 145.4dB DL, 140.7dB UL (20 Kbps, FDD) | 155.7dB | Deep coverage: 164dB +3 | 155.7dB (at 23dBm) | Deep coverage: 164dB |
| Data rates (peak) | DL: 10 Mbps, UL: 5 Mbps | ~800 Kbps (FD-FDD) 300/375 Kbps DL/UL (HD-FDD) | 30kbps (HD-FDD) | DL/ UL: 4 Mbps FD-FDD@5MHz | TBS in 80/ 105Kbps 1352/ 1800 peak rates t.b.d. |
| Latency | Legacy LTE: < 1s | ~ 5s at 155dB | <10s at 164 dB | At least the same as Cat. M1 Legacy LTE (normal MCL) | At least the same as Cat. NB1, some improvements are FFS |
| Mobility | Legacy support | Legacy support | Cell selection, re-selection only | Legacy support | More mobility compared to Cat. NB1 |
| Positioning | Legacy support | Partial support | Partial support | OTDA with legacy PRS and Frequency hopping | 50m H target, new PRS introduced. details FSS. UTDOA under study |
| Voice | Yes (possible) | No | No | Yes | No |
| Optimizations | n/a | MPDCCH structure, Frequency hopping, repetitions | NPDCCH, NPSS/NSSS, NPDSCH, NPUSCH, NPRACH etc., frequency hopping, repetitions, MCO | Higher bandwidth will be DCI or RRC configured, Multi-cast e.g. SC-PTM | Multi-cast e.g. SC-PTM |
| Power saving | DRX | eDRX, PSM | eDRX, PSM | eDRX, PSM | [eDRX, PSM] |
| UE complexity BB | 100% | ~45% | < 25% | [~55%] | [~25%] |

## 3. IOT DEPLOYMENT STRATEGY

The IoT evolution is different from the regular mobile broadband evolution, where the latter is focusing on the connectivity only while the IoT evolution should be addressed from an end-to-end prospective. The IoT connectivity is only 5% to 10% of the IoT value chain, and therefore, service providers should focus on the end-to-end use case. From a customer perspective, the most important factor is the value of the IoT use case and the associated business case (BC) behind it. Therefore, the focus on connectivity has been driven by the legacy telecom vendor as the main factor, while this is not the case from a commercial point of view. There are a lot of successful and innovative IoT use cases which can be implemented, but they offer a marginal BC. Thus, the service providers should adopt a digital transformation strategy towards providing end-to-end IoT use cases instead of focusing on the connectivity only. The focus on connectivity is in the DNA of the legacy mobile operators, which is not valid any more for the current digital transformation, and the connectivity portion is a small fraction of the IoT/ICT use case.

The legacy infrastructure vendors are still focusing on the broadband dump pipe expansion which is just a capacity expansion that will be deployed based on network traffic growth. Therefore, telecom service providers should adopt digital transformation strategies and focus on monetizing the connectivity and data by offering OTT applications and IoT/ICT use cases. Unfortunately, the legacy telecom vendors are struggling to adopt such digital transformations, and they are still promoting the connectivity as the key factor. However, this is not the case with current decline in voice and data revenue. Therefore, service providers should change their operating models and address the enterprise market using a different approach. The majority of IoT use cases are targeting enterprise customers, and therefore, a new model is mandatory to customize the use case to address the enterprise customer requirement. More importantly, the positive BC is a real challenge for the IoT use cases, especially for the enterprise market.[5]

A simplified typical IoT topology that aligned with ETSI/IEEE and other standardization bodies is illustrated in Figure 3.[5] The horizontal IoT platform will support exiting technologies, including 2G/3G/LTE/WiFi/Fixed and low power wide area (LPWA) networks such as 3GPP, standardized technologies such as LTE-MTC (Rel 13/14) and 3GPP NB-IoT (Rel 13), and even the unlicensed LPWA technologies such as LoRa and Sigfox. The IoT Platform horizontal architecture approach ensures an Inter-operability framework for the diverse set of Gateways, and Sensors across a different and variant set of applications and network connectivities.[5,6]

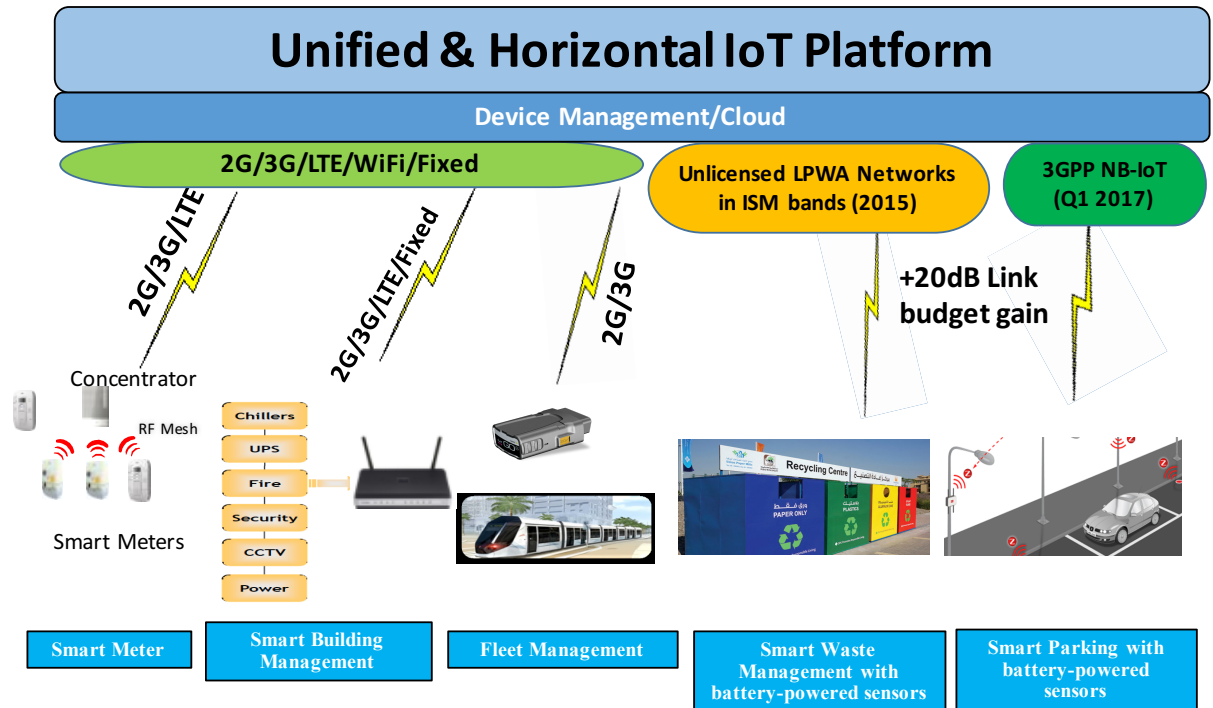

Figure 3: Horizontal IoT Topology Architecture

The horizontal IoT platform will enable customers to *rapidly* build, evolve, and operate scalable and secure IoT solutions for any thing, any data, any network, and any application. A purpose-built and horizontal IoT platform is needed to build and customize IoT applications. These applications can be fully integrated into the network and business processes. Such a solution is secure, scalable, and flexible. Enterprise customers can enjoy enhanced IoT services without the additional burden of managing complex IT infrastructures, applications, and connectivity issues. The IoT platform is a horizontal platform, which factorizes reusable services across different domains, such as device management, data abstraction, protocol translation, data storage and queries, business rules, multi-tenancy, and many others. The IoT platform connects with any device through standard protocols, through an embedded agent, or by using adapters. Once the device is connected and on-boarded, the platform starts to collect data, process and transform this data, and feed applications and integration connectors.[5]

In addition, the IoT platform should be device/gateway agnostic; it is possible to be hosted within premises or at a public or private cloud. It can be integrated with a choice of M2M connectivity platforms for mobile SIM provisioning and management, tariff management, and connectivity management. However, it should be noted that the IoT platform is completely different from the M2M platform, designed mainly for SIM and connectivity management. The IoT platform enables a broad range of public and privately managed IoT services, including provisions for services such as dynamic security, healthcare, hospitality, education, retail, transportation, infrastructure, and financial applications. Delivery of critical real time data, from diverse urban assets to decision makers, business stakeholders, and citizens, enables significant new operational efficiencies, more effective services, and reduced operating costs for cities and enterprises. The IoT platform should be scalable and based on a pay as you grow model. The IoT platform license model is an OPEX model based on the number of connected things. The devices can be classified into light things such as environmental sensors, medium things such as fleet management, and heavy things such as smart building GW with multiple input such as CCTV, fire alarm panels, HAVC, and so on.

To understand the licensing model, let us assume an IoT platform that will manage 1M light sensors, 500k medium things, and 100k heavy things. Assuming 0.5 USD for light thing license, 1 USD for medium thing, and 3 USD for heavy thing, then total annual license will be 1M x 0.5 + 500k x 1 + 3x 100k = 1.6M USD per annum. This model is for illustration purposes only. A unified global license per thing is also a recommended option to avoid capacity bottleneck in the future. A one-time installation and integration fees may be required as well. The IoT platform should be multitenant to offer access to different enterprises use cases.

The classification of the IoT market, along with appropriate technologies, are demonstrated in Figure 4. A typical service provider strategy for IoT evolution is summarized as follows:

- Offer multiple/hybrid technologies per the use case/applications requirements, including throughput, coverage, power, latency, cost, and spectrum.
- The existing networks (2G/3G/LTE/WiFi) can meet the applications that need long range and high data rates. (**10% of IoT market volume.**)
- Short range technologies, such as ZigBee, RF Mesh (802.15.4), PLC, WiFi, etc., can be used for short range applications such as smart meter, smart home, smart parking, etc. (**30% of IoT market volume.**) The short-range devices will be aggregated using a GW that support 2G/3G/LTE to integrate with the IoT platform at a private or public cloud.
- Introducing NB-IoT and/or LTE-M as the mainstream technology for LPWA network use cases. This will be the recommended connectivity for critical IoT applications. (**60% of IoT market volume.**)
- Adopt a hybrid approach for IoT networks connectivity short range, long range, and LPWA networks. This will allow the operators to address a major part of the IoT/M2M verticals and compete with the unlicensed LPWA technologies.
- Nevertheless, a unified/horizontal IoT/M2M platform will be able to manage all technologies

and devices from different networks, different GWs, and different standards and protocols.
- In some specific use cases, such as smart meter, it will be difficult initially to use a horizontal IoT platform, and it is preferable to deploy a specific AMI head-end system.

It is to be noted that horizontal IoT approach is the long-term strategy for service providers. However, some specific vector may need a specific and customized platform. The deployment of a horizontal IoT platform will need to have in-house SW developers to develop and customize applications on top of the IoT platform. This strategy is being adopted by major service providers. This will eliminate the silos verticals and will also mitigate the monopoly of small vendors in specific IoT verticals. However, this strategy should be deployed gradually and smartly to capture all market opportunities [5].

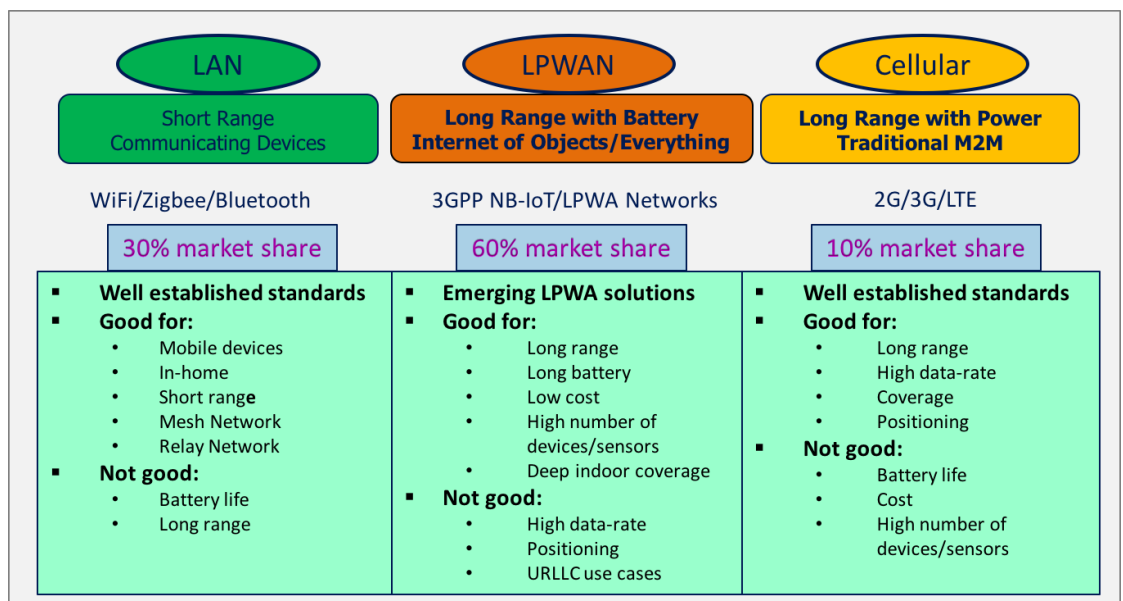

Figure 4: IoT Market Classifications